\documentclass[3p]{elsarticle}

\usepackage{graphicx, AASTeX}
\usepackage{amssymb}
\usepackage{subfig}

\biboptions{comma, sort, authoryear}

\journal{PSS}

\begin{document}

\begin{frontmatter}


\title{Potential Effects of Atmospheric Collapse on Martian Heat Flow and Application to the InSight Measurements}



\author[Stir]{N.~Attree}
\author[OU]{N.~Patel}
\author[Stir,OU]{A.~Hagermann}
\author[DLR]{M.~Grott}
\author[DLR]{T.~Spohn}
\author[Dead]{M.~Siegler}

\address[Stir]{Earth and Planetary Observation Centre, Faculty of Natural Sciences, University of Stirling, UK. Email: {n.o.attree@stir.ac.uk}}
\address[OU]{School of Physical Sciences, The Open University, Milton Keynes, UK}
\address[DLR]{DLR Institute for Planetary Research, Berlin, Germany}
\address[Dead]{Deadman College of Humanities and Sciences, Southern Methodist University, Dallas, USA}

\begin{abstract}

Heat flow is an important constraint on planetary formation and evolution. It has been suggested that Martian obliquity cycles might cause periodic collapses in  atmospheric pressure, leading to corresponding decreases in regolith thermal conductivity (which is controlled by gas in the pore spaces). Geothermal heat would then build up in the subsurface, potentially affecting present--day heat flow --- and thus the measurements made by a heat--flow probe such as the InSight HP$^{3}$ instrument. To gauge the order of magnitude of this effect, we model the diffusion of a putative heat pulse caused by thermal conductivity changes with a simple numerical scheme and compare it to the heat--flow perturbations caused by other effects.  We find that an atmospheric collapse to 300 Pa in the last 40 kyr would lead to a present--day heat flow that is up to $2-8\%$ larger than the average geothermal background. Considering the InSight mission with expected $5-15\%$ error bars on the HP$^{3}$ measurement, this perturbation would only be significant in the best-case scenario of full instrument deployment, completed measurement campaign, and a well--modelled surface configuration. The prospects for detecting long-term climate perturbations via spacecraft heat--flow experiments remain challenging.
\end{abstract}


\end{frontmatter}

\section{Introduction}
\label{S:1}

Terrestrial planets transmit heat from their interiors to their surfaces primarily via convection in the mantle (and sometimes the core) and then conduction through their solid lithospheres. This internal heat is a product of both the initial energy of planet formation, and of radioactive decay of long--lived radioisotopes, and can drive tectonic and volcanic activity as well as planetary magnetic dynamos. Measurements of heat flow through the upper layers of the crust can therefore help to constrain models of a particular planet's formation and composition, in addition to improving our understanding of its lithospheric thickness and crustal geology.

In the case of Mars, various compositional and heat transfer models predict a planetary heat flow of around $20-30$ mW m$^{-2}$ (\citealp{Spohn2018}, and references therein), or just under $20$ mW m$^{-2}$ \citep{Ruiz2011, Parro2017, Egea-Gonzalez2017}, with variations on this number able to differentiate between certain models \citep{Plesa2015}. The main goal of the HP$^{3}$ instrument on the InSight mission is therefore to measure this value for a representative site on Mars, by the use of a self--burrowing probe to record regolith temperatures up to 5 m into the subsurface. This should be deep enough to reach beyond the thermal waves induced by diurnal and annual temperature variations \citep{Spohn2018}, but may be still within the zone directly affected by longer term climate changes. These changes have previously been considered small compared to other perturbations on the global average heat flow, but may contribute a significant source of uncertainty to the HP$^{3}$ measurement that needs to be quantified.

The state of Mars' atmosphere and climate critically depends on the planet's obliquity and orbit \citep{Laskar2004}. It has been shown \citep{WG2007,WG2009,WG2009b,WG2014} that changes in the planet's obliquity and orbit could trigger periodic collapse of the Martian atmosphere, which would lead to dramatic changes in regolith thermal conductivity and, therefore, subsurface heat transport.

During periods of low obliquity, occurring in cycles of roughly 120 kyr and 1.3 Myr periods, polar insolation is reduced, leading atmospheric CO$_{2}$ to condense onto the polar ice caps, thereby causing a dramatic reduction in surface pressure \citep{Manning2006}. Since the thermal conductivity of Martian dry regolith is dominated by gas filling the pore spaces \citep{Presley1997}, this leads to a corresponding decrease in its ability to transfer heat, resulting in a steeper temperature gradient in the regolith, and thus raising temperatures below the affected regolith layer. As atmospheric pressure, and regolith thermal conductivity, increases again, the heat stored underneath this insulating layer will diffuse away and the thermal gradient will return to its long--term equilibrium state. During this time, however, an increased heat flow (relative to average conditions) will be found in the near--surface, which may bias the results of the HP$^{3}$ experiment.

In this work, we therefore seek to model the propagation and decay of a subsurface heat accumulation caused by a temporary decrease in near--surface thermal conductivity, following the atmospheric pressure trends described in the literature. We then quantify the effects of the heat release on the present--day thermal profile, and discuss its potential detectability in (and/or biasing of) the results of a heat--flow experiment such as InSight's HP$^3$. Since the effect is small, we seek to place upper limits on its perturbation of present--day heat flow, i.e.~the largest possible uncertainty on the InSight results due to this effect. The rest of the paper is outlined as follows. In Section \ref{S:2}, we describe in more detail the mechanism of atmospheric collapse and regolith thermal conductivity change, and examine estimates of its magnitude from the literature. In Section \ref{S:3}, we describe a simple numerical model developed to calculate the expected heat--flow perturbation from a given thermal conductivity change. In Section \ref{S:4}, we show the results of the model, validating it against the literature from Section \ref{S:2}, and examining the effects of varying key parameters. This is followed by a discussion, with references to the detectability by a heat--flow probe such as InSight, in Section \ref{S:5}, and our conclusions, in Section \ref{S:6}.

\section{Background}
\label{S:2}
\subsection{Climate changes associated with obliquity cycles}
\label{S:2.1}

\begin{figure}
\centering\includegraphics[width=1.0\linewidth]{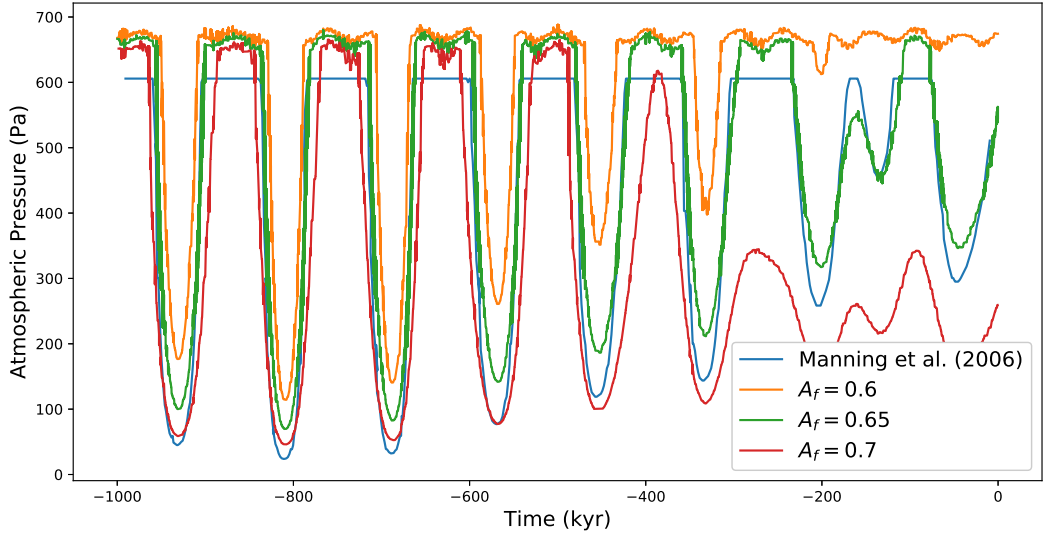}
\caption{Martian atmospheric pressure over the last million years, from climate models by \citet{Manning2006} and \citealp{WG2009b} (for three different values of polar CO$_{2}$ ice albedo, $A_{f}$).}
\label{Plot_Pt}
\end{figure}

The decrease in pressure during atmospheric collapse is found to be controlled by the thermodynamic balance between condensation and sublimation of CO$_{2}$ ice at the poles, which is, in turn, dependent on the ice's albedo \citep{Manning2006, WG2009b}. Higher albedos mean more solar energy is reflected, reducing energy input and allowing more CO$_{2}$ to condense, increasing the magnitude of atmospheric collapse. This can be seen in Figure \ref{Plot_Pt}, which shows atmospheric pressure over the last million years for several different polar albedos ($A_{f}$), as calculated by \citet{Manning2006} and \citet{WG2009b}. Periodic collapses following the 120 kyr obliquity cycle are visible, with a magnitude that is controlled by $A_{f}$ and ranges from the present--day $\sim650$ Pa atmospheric pressure to a minimum of $\sim30$ Pa. Some polar albedo values ($A_{f}\geq0.7$) are effectively ruled out as they cannot reproduce present day pressures, while others ($A_{f}\leq0.6$) do not produce any collapse in the last few hundred thousand years. Most models show a collapse to $\sim300$ Pa around 40-50 kyr ago as the most recent.

In addition to atmospheric pressure changes, varying obliquity will also change average surface temperatures. \citet{Forget13} show that global mean temperature is inversely correlated with obliquity, but that feedback effects (such as growth and retreat of the polar caps, with their above mentioned high albedo) mean the variation is small ($\lesssim5$ K) between obliquities of $\sim15-35^{\circ}$, as occurred over the past Myr \citep{Laskar2004}. Local surface temperatures can vary significantly more than the global average of course, but \citet{Haberle2003} again show a relatively small $\sim\pm2~\mathrm{K}$ variation in equatorial annual mean surface temperature over this obliquity range.

\subsection{Regolith response to changes in atmospheric pressure}

\begin{figure}
\centering\includegraphics[width=1.0\linewidth]{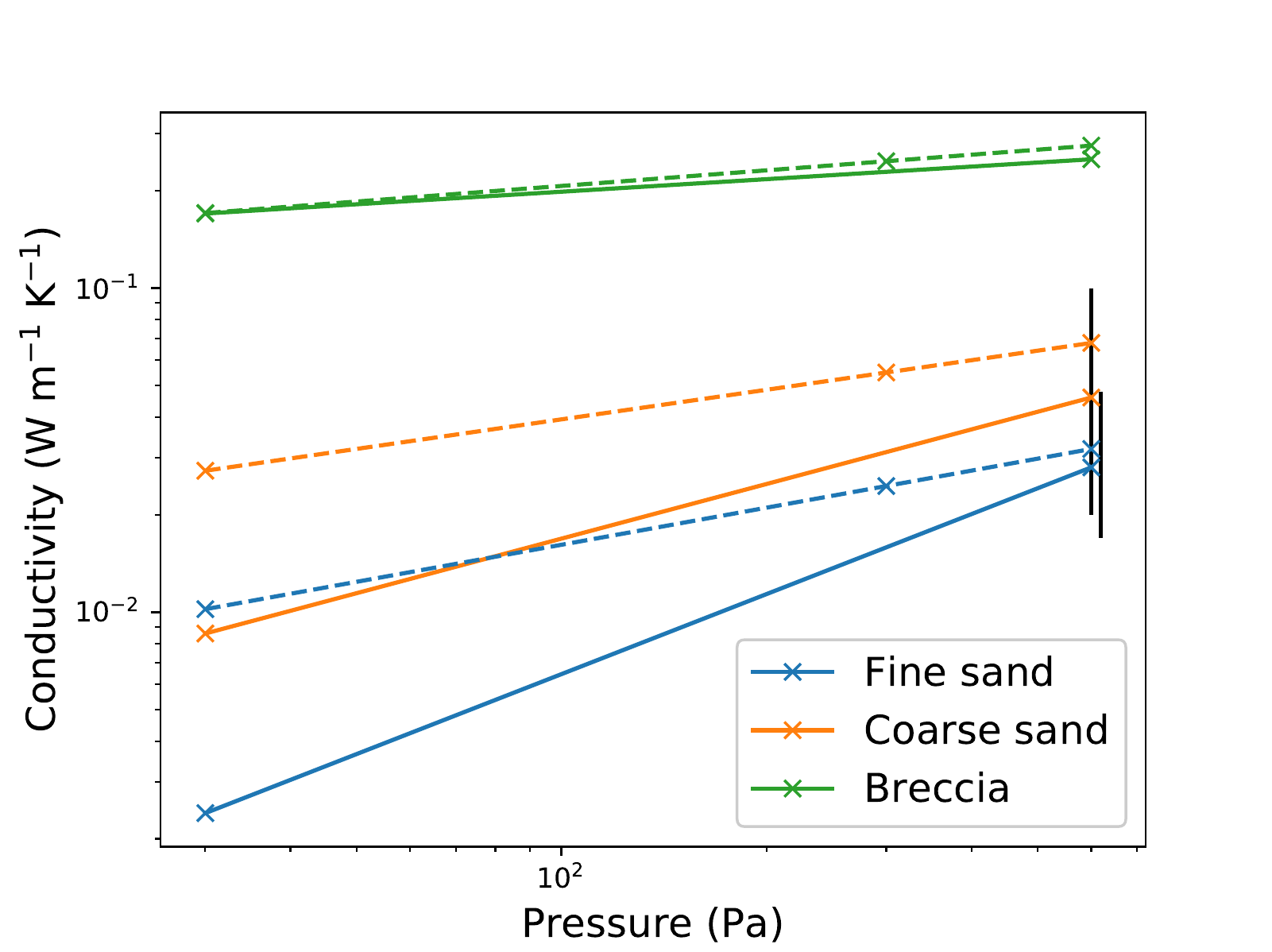}
\caption{Pressure dependence of thermal conductivity from \citet{WG2009} (solid lines) and \citet{Presley1997} (dashed) for three different materials: fine sand of 100 $\mu$m particle diameter, coarse sand of 500 $\mu$m, and breccia of 1 cm size. \citet{WG2009} values are taken from their Table 1 for 30 Pa and 600 Pa pressure, \citet{Presley1997} values are calculated for 30, 300 and 600 Pa from their Eqn.~1. Vertical bars show the present--day thermal conductivity range expected for surface material at the InSight landing site from \citet{Grott} and \citet{Morgan2017} (offset for clarity).}
\label{Plot_kP}
\end{figure}

CO$_{2}$ gas filling the regolith pores will respond quickly compared to the changing surface pressure curves of Figure \ref{Plot_Pt}. \citet{Lorenz2015}, for example, calculate, based on experimentally measured diffusion coefficients, that seasonal $\sim200$ Pa pressure changes can propagate down to several hundred metres regolith depth within a single Martian year. Experimental work by \citet{Presley1997} has then found that regolith thermal conductivity $k$ is dominated by heat transport by the pore gas, and to empirically show a roughly power--law dependence on pressure (with an exponent of 0.6). \citet{WG2007} extended this with a more complicated first--principles model based on various parameters of porous materials (e.g.~pore size, packing fraction and material bulk thermal conductivity). The behaviour of both models is similar for relevant Martian pressures, as shown by \citet{WG2007}. \citet{WG2009} list conductivities for fine-grained (100 $\mu$m particle diameter) and coarse-grained sand (500 $\mu$m particle diameter) as well as breccia (with a 1 cm grain size) at 600 and 30 Pa atmospheric pressures. Figure \ref{Plot_kP} shows these values, alongside conductivities calculated at the same particle sizes with Eqn.~1 of \citet{Presley1997}. Present--day regolith conductivities fall within the range given by \citet{Grott} and \citet{Morgan2017} for the material at the InSight landing site. Regolith conductivities during a severe atmospheric collapse are around five times smaller than present--day values for coarse sand and ten times smaller for fine sand. 

\subsection{Resulting heat--flow perturbation}

\citet{WG2009,WG2009b} coupled their regolith model with a climate model, using the time--varying atmospheric pressure of \citealp{Manning2006} (blue curve in Fig.~\ref{Plot_Pt}) to explore subsurface temperature changes over the last Myr of Martian history at a range of latitude locations and subsurface profiles (such as, typically 100 m of fine or coarse-grained sand overlaying breccia and then solid rock).

In all cases, they found temperatures perturbed from the pre-collapse average, with a maximum perturbation occurring at the regolith/breccia boundary at depth $z_{0}$, explained by the thermal conductivity mismatch here when the regolith's thermal conductivity is drastically reduced, and the breccia's only slightly, during atmospheric collapse. Thus, the amount of geothermal heat escaping through the regolith is reduced and energy is stored below this more insulating layer, increasing temperatures. Over time, the temperature below the regolith keeps increasing until a new equilibrium state is reached that reflects the changed ratio of thermal conductivities above and below this boundary. However, long before this equilibrium can be reached, the atmospheric pressure is restored (essentially instantaneously, compared to the temperature response), regolith thermal conductivity increases again, and the subsurface temperature profile begins to return to the old, flatter profile. The increased temperature centred on $z_{0}$ diffuses and an increased heat flow is found until the background thermal balance is restored.

The temperature perturbation, relative to the background temperature structure, is a function of depth and time since the end of collapse, $\Delta T (z, t)$. \citet{WG2009} show $\Delta T (z, t)$ for a number of different models and times, noting that the perturbation is largest (around 30 K at the regolith/breccia boundary) just after the end of a collapse, and decays away towards the present day, while its magnitude is largely independent of latitude (and, therefore, of surface temperature trends). Typically the perturbation is only a few K at the present day.  

We can calculate the transient excess heat flow associated with this perturbation by \citet{Grott}:
\begin{equation}
\Delta F(z, t) = k\frac{\partial}{\partial z}\Delta T(z, t),
\label{FluxChange}
\end{equation}
where $k$ is the present--day material thermal conductivity (since we are interested in the period after it has `reset' to present--day atmospheric pressure values). Taking some examples from \citet{WG2009}, present--day gradients of around $0.02-0.03$ K m$^{-1}$ are seen in the perturbation heat flow at 10$^{\circ}$ north latitude in the near surface for both coarse and fine-grained sand (i.e.~$\Delta T(z_{0-5 m}, t_{now})\sim0.02-0.03$ K m$^{-1}$). Using Eqn.~\ref{FluxChange} with a present--day regolith thermal conductivity of $k=0.02-0.1$ W m$^{-1}$ K$^{-1}$, as given by \citet{Grott}, corresponds to perturbations of $0.4-3.0$ mW m$^{-1}$, i.e.~less than $10\%$ of the assumed $30$ mW m$^{-2}$ average geothermal heat flux.

In this study we aim to investigate the effect of such a temperature excursion on present--day heat--flow measurements such as InSight's HP$^{3}$. According to \citet{Spohn2018} and \citet{Grott2019}, the accuracy of the HP$^{3}$ instrument, when taking into account the known sources of uncertainty, should be $5-15\%$. Perturbations associated with atmospheric collapse would therefore be around the level of the existing noise, and their detection would appear challenging. Nevertheless, they may be have an important influence on the InSight results and might otherwise bias the final derived geothermal flux value, if their effects are not taken into account. We thus wish to investigate atmospheric collapse in more detail, noting that the work of \citet{WG2009} was for a latitudinal location and regolith depth different to that of InSight's landing location. Below, we present a simple numerical model of the propagation of a thermal wave, associated with near--surface thermal conductivity changes, and examine the effects of various parameters on its perturbation to present--day heat flow. We consider this separately from the direct effects of the climate change itself, in order to quantify a strongest--case scenario for atmospheric collapse, to examine its detectability and whether it warrants detailed study with respect to InSight.

\section{Method}
\label{S:3}

We use a fully implicit, finite control volume approach (see, e.g. \citealp{Patankar}) to solve the one dimensional, time-dependent heat diffusion equation

\begin{equation}
\rho(z) c(z)\frac{\partial T}{\partial t} = \frac{\partial}{\partial z}\left(k(z)\frac{\partial T}{\partial z}\right).
\label{1D}
\end{equation}
Here $\rho$ and $c$ are the bulk density and specific heat, respectively, and temperature $T=T(z,t)$ is a function of both time and depth.

We define a simple three-layer structure, to represent regolith with a depth of $z_{0}$ with breccia/megaregolith below and bedrock below that. Regolith depth is expected to vary across the surface of Mars. In order for our results to be applicable to InSight, we need to take into account that unconsolidated surface material is thought to extend only $3-17$ m deep \citep{Warner2017, Golombek2017} at the landing site, and so we investigate the effects of varying $z_{0}$ between 5 and 10 m, with a fixed 10 m thick breccia layer between this and the bedrock, which itself extends to 2 km depth.

 Within each layer, the thermal conductivity is fixed at a single value, and the density and specific heat for both the regolith and breccia are set to $\rho=1750$ kg m$^{-3}$ and $c=600$ J kg$^{-1}$ K$^{-1}$, respectively, as in \citet{Grott}. Following \citet{Paton16}, the bedrock is set to $\rho=2600$ kg m$^{-3}$ and $c=800$ J kg$^{-1}$ K$^{-1}$. In reality, thermal conductivity should be a function of time and depth, $k=k(z,t)$, dependent not just on the layering, but variations within the layers. Thermal conductivity in the near-surface regolith layer should increase with depth (due to decreasing porosity and/or a low $k$ surface dust layer), but, according to standard models (e.g.~\citealp{Grott}), quickly approaches a constant value. Thus, below a few metres depth and within the regolith layer, thermal conductivity should be approximately constant, and a model with a constant value is therefore a reasonable approximation at the depths ($\geq5$ m) and latitude (near equatorial, where ground ice is negligible) of interest to InSight. As noted above, the temperature perturbation effect comes from the difference between conductivities at layer boundaries, so by having sharp changes, rather than smoothly varying thermal conductivity with depth, we ensure that our model produces a maximum, strongest--case effect.
 
Since many of the parameters in \citeauthor{WG2009}'s model are largely unknown (e.g.~polar albedo, regolith particle size, packing fraction and pore size, etc.), and we are interested in finding the maximum, worst-case scenario for the effects of atmospheric collapse, we follow several further approximations.

Firstly, we take a constant surface temperature (equal to the present--day seasonal average of 220 K at 4$^{\circ}$N) as the top boundary condition (the bottom boundary condition is taken as a constant geothermal heat flux, $F=20$ mW m$^{-2}$, as in \citealp{Grott}). A constant surface temperature is used, rather than a time-varying climate model, to investigate purely the effects of thermal conductivity changes as distinct from the direct temperature changes associated with the obliquity cycles. This is justified by the small variations in mean annual surface temperature, as noted in section \ref{S:2.1}, as well as the swift decay with time of their effects, as noted by \citeauthor{Grott}, 
and the minimal effect on subsurface perturbations the surface temperature is noted to have by \citeauthor{WG2009b}.

Secondly, we simplify the collapse itself to avoid the complication of the many unknown parameters. We set the thermal conductivities to those appropriate for some pressure P, followed by a return to present--day values after a set duration (essentially taking a square well pressure--time dependence, rather than choosing one of the curves of Fig.~\ref{Plot_Pt}). This means we get the maximum possible temperature increase at $z_0$, as thermal conductivity is always at its minimum during the entire duration of the collapse.

The procedure is as follows: the model is run for several thousand years with present--day thermal conductivities, to reach an equilibrium temperature profile. Conductivities are decreased to their atmospheric collapse values for some duration and then returned to their present values. The perturbed temperature profile, relative to the pre--collapse equilibrium, is measured immediately after the end of the collapse and for up to 1 Myr subsequently, in order to observe its decay towards the background temperature profile. We vary the duration of the collapse, as well as the depth of the regolith/breccia boundary, their conductivities, and the minimum atmospheric pressure, to investigate their effects on the subsurface temperature.

\section{Results}
\label{S:4}

We first validate our model by reproducing the results of \citet{WG2009}. Using a $z_{0}=100$ m deep regolith layer of coarse sand over breccia and rock (conductivities of $k=0.046, 0.25$ and 2 W m$^{-1}$ K$^{-1}$, respectively) and a 30 kyr collapse to 30 Pa (collapse conductivities of $k=0.0086, 0.17$ and 2 W m$^{-1}$ K$^{-1}$, respectively), we find a maximum $\sim30$ K temperature perturbation at $z_{0}$, that decays to a few K at present day, matching the results of \citet{WG2009}.

\begin{figure}
\centering\includegraphics[width=1.0\linewidth]{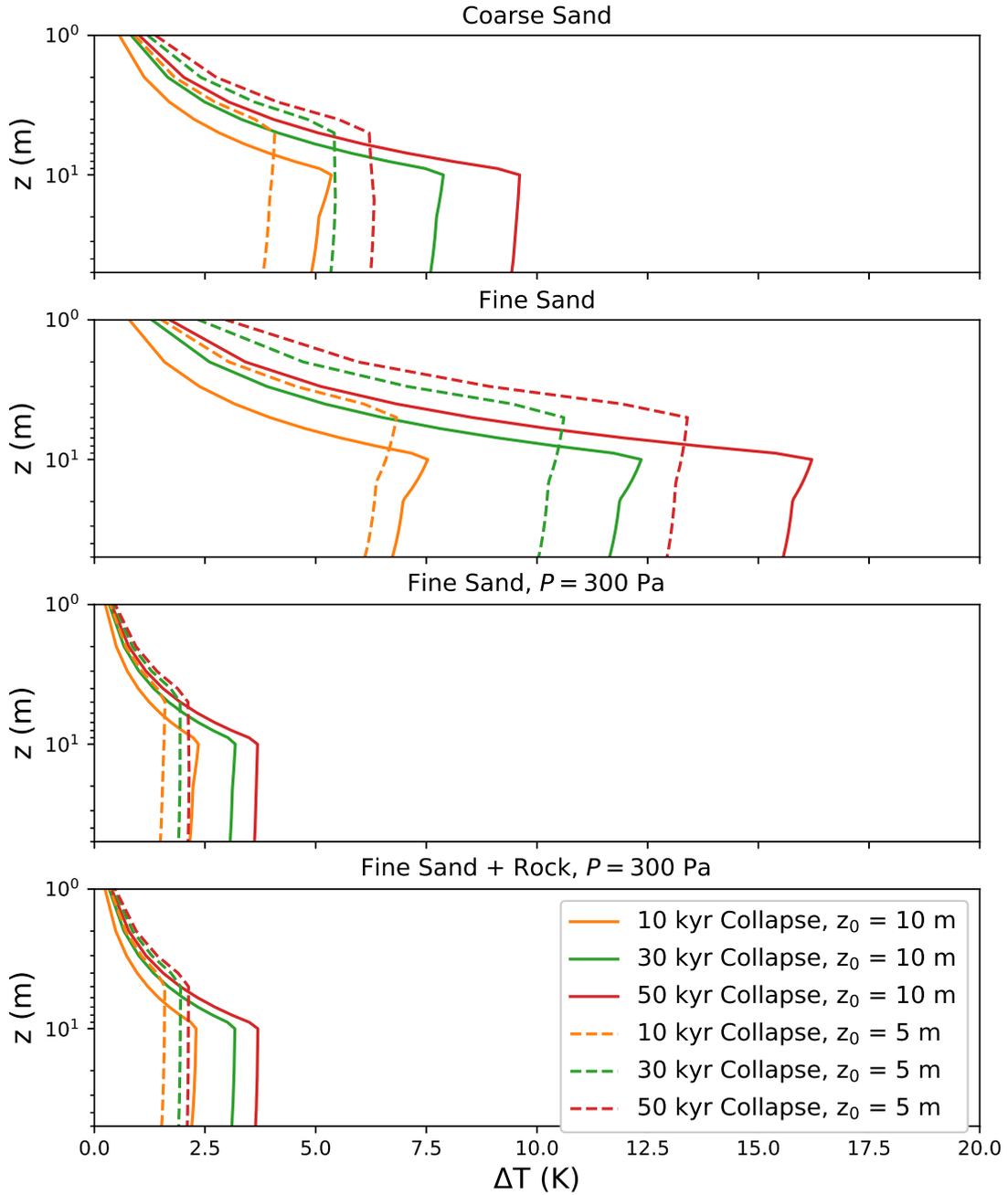}
\caption{Temperature perturbations expected at the end of an atmospheric collapse from various periods of reduced regolith thermal conductivity with varying regolith depth $z_{0}$ (solid: $z_{0}=10$ m, dashed: $z_{0}=5$ m). The scenarios are, from top: collapse to 30 Pa with coarse--grained regolith, collapse to 30 Pa with fine--grained regolith, collapse to 300 Pa with fine--grained regolith, and collapse to 300 Pa with fine--grained regolith overlaying bedrock.}
\label{Plot_Figure3}
\end{figure}

Next, we consider the effect of the heat--flow perturbation on the InSight measurements by selecting a subsurface structure more appropriate for the InSight landing site. Figure \ref{Plot_Figure3} (first row) shows the temperature perturbation, $\Delta T (z, t)$, at the end of collapse for three different collapse durations, and regolith depths of 5 and 10 m of coarse sand. Thinning the insulating regolith layer reduces the proportion of geothermal heat retained below, resulting in much smaller temperature perturbations than given by \citet{WG2009}. The maximum perturbation is now around 10 K.

\begin{figure}
\centering\includegraphics[width=1.0\linewidth]{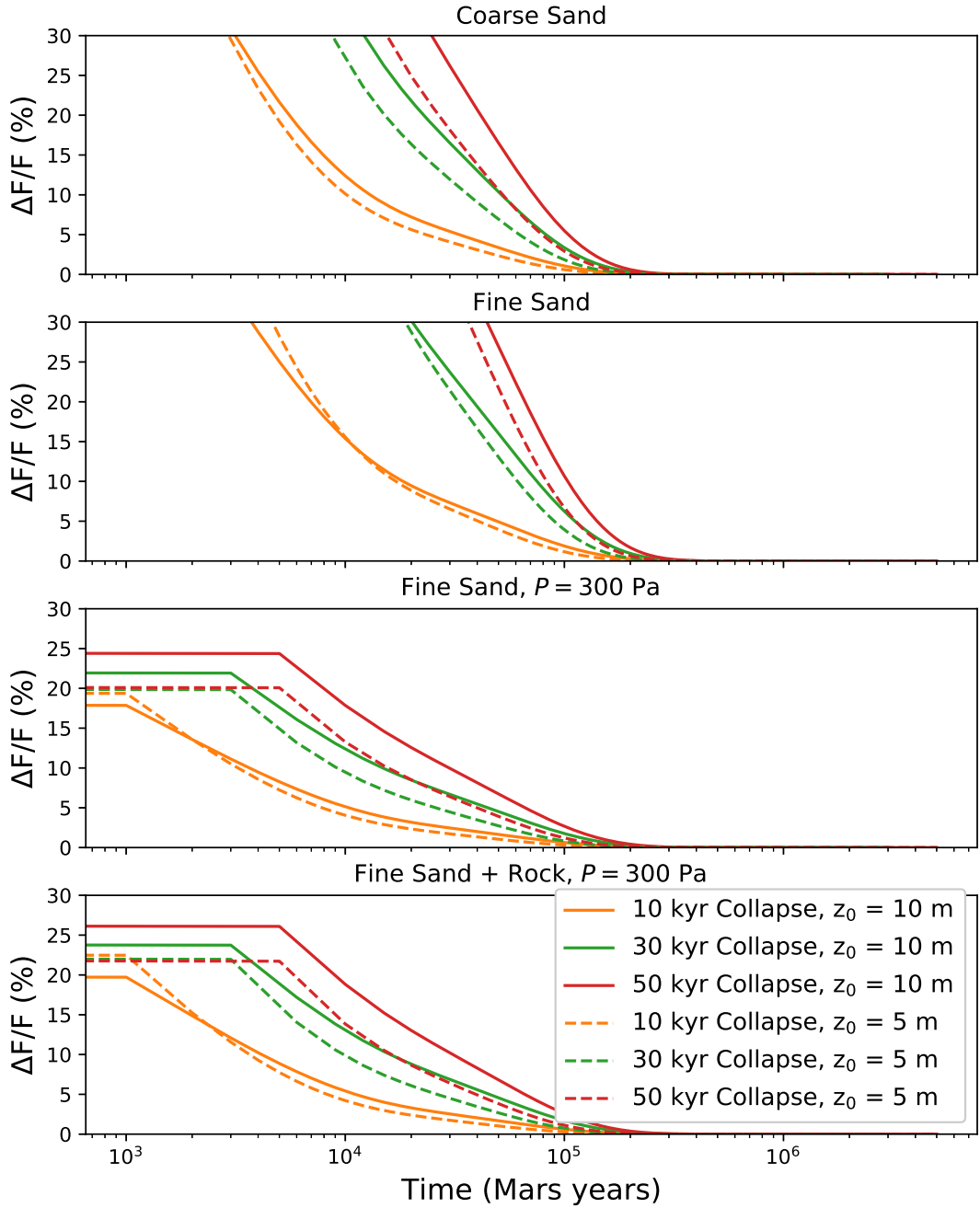}
\caption{Heat--flow perturbations, with time since the end of an atmospheric collapse, expected at 5 m depth (solid: $z_{0}=10$ m, dashed: $z_{0}=5$ m). The scenarios are, from top: collapse to 30 Pa with coarse--grained regolith, collapse to 30 Pa with fine--grained regolith, collapse to 300 Pa with fine--grained regolith, and collapse to 300 Pa with fine--grained regolith overlaying bedrock.}
\label{Plot_model}
\end{figure}

When we now advance our simple thermal model forwards in time, after the end of the atmospheric collapse, the heat pulse is observed to decay back to background conditions. At each time and depth we can calculate the excess heat flow which would be seen, above the average geothermal flux, by subtracting the background profile away to find $\Delta T(z, t)$,
and then using Eqn.~\ref{FluxChange}, with present--day conductivities. Figure \ref{Plot_model} (first row) shows the heat--flow perturbation with time since the end of the collapse, at the InSight measurement depth ($z=5$ m), for each of the regolith depths and collapse durations from above. In all cases the initially large perturbation is negligible after a few hundred thousand years. The nominal $z_{0}=10$ m, 30 kyr perturbation decays away to around $15\%$ of the assumed $20$ mW m$^{-2}$ heat flux after $\sim40$ kyr, while the effect is larger/smaller for longer/shorter collapse durations and thicker/thinner regolith. 

As discussed above, the temperature perturbation arises at the boundary between two layers of differing conductivities, so what happens if this difference is greater? We replace the coarse sand conductivities with the smaller values for fine sand from \citealp{WG2009} ($k=0.028$ and 0.0024 W m$^{-1}$ K$^{-1}$ for 600 and 30 Pa, respectively), keeping the breccia and bedrock the same, and repeat the experiment. The second rows of Figures \ref{Plot_Figure3} and \ref{Plot_model} show the resulting temperature and heat--flow perturbations, respectively. Sure enough, the smaller regolith conductivities, relative to the unchanged breccia values, means that the fine sand acts as a better insulator, trapping more geothermal heat during collapse. Upon return to present pressures, the large temperature perturbation takes longer to diffuse away, leading to increased heat--flow perturbations, i.e.~over $20\%$ of present--day heat flow for the 30 kyr case after 40 kyr, and nearly $40\%$ for a 50 kyr collapse!

%\begin{figure}[h]
%\centering\includegraphics[width=1.0\linewidth]{Figure3_fine.eps}
%\caption{Temperature perturbations expected at the end of a collapse to 30 pa from various %periods of reduced regolith thermal conductivity with varying regolith depth $z_{0}$ %(solid: $z_{0}=10$ m, dashed: $z_{0}=5$ m) for fine-grained regolith.}
%\label{Plot_Figure3_fine}
%\end{figure}

%\begin{figure}[h]
%\centering\includegraphics[width=1.0\linewidth]{Plot_FluxChange_model_fine.eps}
%\caption{Heat--flow perturbation, with time since the end of an atmospheric collapse to 30 %pa, expected at 5 m depth (solid: $z_{0}=10$ m, dashed: $z_{0}=5$ m) for fine-grained regolith.}
%\label{Plot_model_fine}
%\end{figure}

The above would be significant perturbations and should be detectable by HP$^3$ on InSight. However, noting Fig.~\ref{Plot_Pt}, the chance of a recent collapse reaching 30 Pa atmospheric pressures appear very remote. The models suggest that recent collapses, in the last few hundred thousand years, only reach a minimum pressure of around 300 Pa. We therefore investigate such a collapse by using the present--day thermal conductivity of fine sand from \citet{WG2009}, and interpolating a value at 300 Pa pressure from the log-log plot (Fig.~\ref{Plot_kP}) of $k=0.016$ W m$^{-1}$ K$^{-1}$. We do this, rather than using the \citet{Presley1997} results shown in the same plot, both for consistency with our previous models, and because, as can be seen in the plot, the \citet{WG2009} conductivities decay fastest with pressure, again giving us a strongest--case effect.

%\begin{figure}[h]
%\centering\includegraphics[width=1.0\linewidth]{Figure3_fine300Pa.eps}
%\caption{Temperature perturbations expected at the end of a collapse to 300 Pa from various periods of reduced regolith thermal conductivity with varying regolith depth %$z_{0}$ (solid: $z_{0}=10$ m, dashed: $z_{0}=5$ m) for fine-grained regolith.}
%\label{Plot_Figure3_fine300Pa}
%\end{figure}

%\begin{figure}[h]
%\centering\includegraphics[width=1.0\linewidth]{Plot_FluxChange_model_fine300Pa.eps}
%\caption{Heat--flow perturbation, with time since the end of an atmospheric collapse to 300 pa, expected at 5 m depth (solid: $z_{0}=10$ m, dashed: $z_{0}=5$ m) for fine-grained regolith.}
%\label{Plot_model_fine300Pa}
%\end{figure}

Row three of Figures \ref{Plot_Figure3} and \ref{Plot_model} show the resulting temperature and heat--flow perturbations. As expected, temperature perturbations are now much reduced, not exceeding 5 K in any of the models. This results in smaller heat--flow perturbations of between $\sim2-8\%$ and around $5.5\%$ for a 30 kyr collapse and 10 m of regolith. Perturbations will be even smaller for coarser grained material.

Finally, we can ask what the perturbations would look like in a situation where the fine-grained regolith layer overlays solid bedrock directly, with no intermediate breccia/megaregolith layer. The large thermal conductivity mismatch at $z_{0}$ might be expected to produce a large temperature perturbation here, but as shown in the bottom row of Figures \ref{Plot_Figure3} and \ref{Plot_model}, the effects are similar to the above three-layered cases, with an only slightly larger perturbation. This demonstrates that it is the properties of the regolith (i.e.~its thickness, present--day thermal conductivity and corresponding pressure dependence), and the collapse itself (duration and minimum pressure reached), that are the dominant factors in determining the heat--flow perturbation associated with atmospheric collapse.

%\begin{figure}[h]
%\centering\includegraphics[width=1.0\linewidth]{Figure3_finerock300Pa.eps}
%\caption{Temperature perturbations expected at the end of a collapse to 300 Pa from various periods of reduced regolith thermal conductivity with varying regolith depth $z_{0}$ (solid: $z_{0}=10$ m, dashed: $z_{0}=5$ m) for fine-grained regolith over bedrock.}
%\label{Plot_Figure3_finerock300Pa}
%\end{figure}

%\begin{figure}[h]
%\centering\includegraphics[width=1.0\linewidth]{Plot_FluxChange_model_finerock300Pa.eps}
%\caption{Heat--flow perturbation, with time since the end of an atmospheric collapse to 300 pa, expected at 5 m depth (solid: $z_{0}=10$ m, dashed: $z_{0}=5$ m) for fine-grained %regolith over bedrock.}
%\label{Plot_model_finerock300Pa}
%\end{figure}

\section{Discussion}
\label{S:5}

We have presented upper limits for the ability of thermal conductivity changes due to atmospheric collapse to perturb present--day heat flow on Mars. In particular, these are upper limits because regolith thermal conductivity should decrease near the surface, reducing heat--flow perturbations here. Although our model of an instantaneous change in thermal conductivity is relatively simple, the conclusions drawn from it are relatively robust. Thermal conductivity changes which smoothly follow the pressure curves of Fig.~\ref{Plot_Pt}, rather than changing instantaneously as in our model, would lead to a regolith that spends less time as an effective insulating layer, reducing heat build up. Additionally, smoothly varying regolith size/thermal conductivity with depth would also reduce the thermal conductivity difference at the layer boundaries, again resulting in smaller perturbations than presented here.

We thus conclude that our estimated heat--flow perturbation, of around $\sim2-8\%$ for an atmospheric collapse to 300 Pa in the last 40 kyr, is a reasonable upper limit. Coarser grained material, towards the higher end of the expected thermal conductivity range, would lead to even smaller perturbations. This is the positive excess we would expect a present--day heat--flow experiment such as InSight's HP$^3$ to measure, on--top of an average geothermal flux of $F=20$ mW m$^{-2}$. If a severe atmospheric collapse to 30 Pa occurred around 40 kyr ago, the perturbations could be larger, up to $\sim20-40\%$, but this is considered unlikely \citep{Manning2006, WG2007}. In relation to the InSight measurements, it is instructive to compare this heat--flow perturbation to those estimated for other sources of uncertainty considered.

\citet{Spohn2018} break down the error budget from sources relating to the HP$^{3}$ measurement process itself, as well as the external uncertainty which comes from averaging out the annual thermal wave. They estimate a total uncertainty of $2.2$ mW m$^{-2}$ or $10.8\%$ for the nominal case of the instrument reaching 3 m depth. If the probe reaches the full 5 m depth, the $10\%$ of this error coming from of the annual heatwave is reduced and \citeauthor{Spohn2018} estimate a $5\%$ uncertainty. Both these estimates are for 0.6 Martian years worth of measurements. If the full year's worth can be obtained, the annual heatwave error can be even further reduced, but the total instrument error of around $4\%$ is unlikely to be improved upon. Thus, in the relatively optimistic scenario of good instrument deployment and full measurement time, our maximum-strength atmospheric collapse signal may just be detected above the error bars; whereas, in the worst-case scenario, it is likely to be lost in the instrument noise.

Further sources of error are not taken into account by \citeauthor{Spohn2018}. For example, \citet{Grott} estimate the perturbation caused by the direct surface temperature changes due to the obliquity cycles themselves. Modelling the propagation of thermal waves into an homogeneous semi-infinite half space, they estimate perturbations of a few percent, for a 10 K step-change $0.1-1$ Myr before present, and by $<15\%$, for 10 K periodic forcing every 120 kyr. From the discussion in Section \ref{S:2} above, this appears to be too large. Assuming instead periodic forcing with an amplitude of $\pm2$ K in surface temperature, we can use \citeauthor{Grott}'s Eqn.~6 with $k=0.046$ W m$^{-1}$ K$^{-1}$ to estimate a present-day perturbation of $\sim1.5\%$ at 5 m depth. The direct temperature perturbation of the obliquity cycles therefore appears to be similar to, or smaller than, the thermal conductivity change effect, for reasonable parameter choices. 

Additionally, if the recent (centuries before present) Little Ice Age on Earth was caused by solar changes and also affected Mars, it might cause perturbations of $5-10\%$, according to \citet{Lorenz2015}. \citet{Morgan2017}, meanwhile, showed that atmospheric convection into the HP$^{3}$ borehole should be negligible, but that changes in surface temperature from dust redistribution during landing may have an effect. After landing, shadowing by and re--radiation from the lander will also cause temperature changes to propagate into the subsurface. According to thermal modelling by \citet{Siegler2017}, these should be limited to $\Delta F<3$ mW m$^{-2}$, or $15\%$, when measured below 1 m and away from lander body, although complicated seasonal changes may need to be averaged out over time. Finally, local dust storms reduce atmospheric opacity and redistribute surface dust, and can perturb heat flow by $2$ mW m$^{-2}$ to depths of up to $1-2$ m (for $k=0.02-0.05$ W m$^{-1}$ K$^{-1}$), while global dust storms, as occurred in Mars year 28, can cause $1.5-2.5$ mW m$^{-2}$ perturbations as deep as $2-3$ m \citep{Plesa2016}.

All of these additional perturbations are of a similar magnitude to those expected from thermal conductivity changes associated with atmospheric collapse. We note, however, that the presence of a dust storm is easily observed and it should, therefore, be possible to model its effect and recover a good estimate of the undisturbed geothermal flux. This is similarly the case for the lander, re-radiation and shadowing, which can be modelled as shown by \citet{Siegler2017}. Assuming good instrument deployment, full spacecraft functionally over a Martian year and a well--modelled surface configuration, the largest perturbations are therefore those associated with climate change, e.g.~several percent flux change from the direct effect of the obliquity cycles, $\sim2-8\%$ change from atmospheric collapse associated with them, and $5-10\%$ from any recent Little Ice Age.

\begin{figure}
\centering\includegraphics[width=1.0\linewidth]{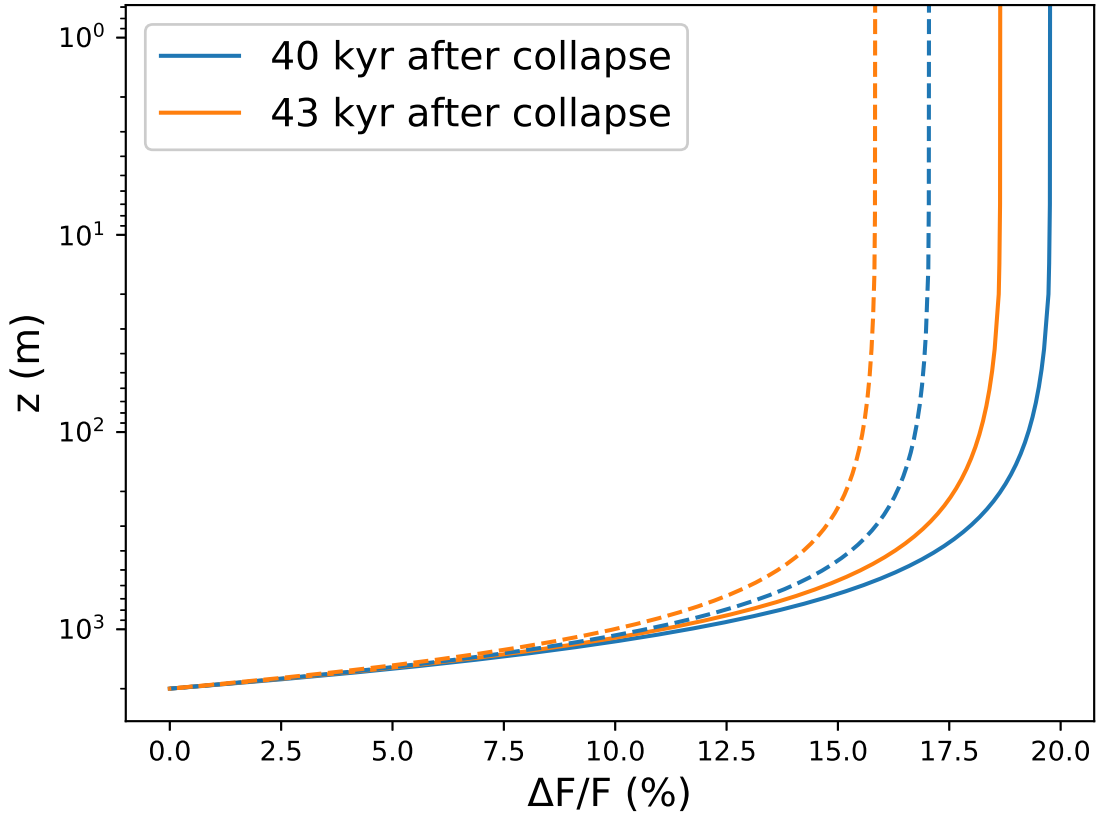}
\caption{Heat-flow perturbation with depth for two times after the end of an extreme atmospheric collapse to 30 Pa, and regolith depths $z_{0} = 10$ m (solid) and $z_{0} = 5$ m (dashed). Gradients in the region of interest are essentially constant, and changes with time negligible for this timeframe.}
\label{Plot_model_withdepth}
\end{figure}

Disentangling these different climate change signals from their magnitudes alone will be difficult. Can we say anything more about the direction of the signals, or how they vary in time and depth? Figure \ref{Plot_model_withdepth} shows the depth profile of the atmospheric collapse heat-flow perturbation at 40 kyr after the end of our nominal 30 kyr collapse (for the strongest-case scenario of fine-grained regolith and a collapse to 30 Pa). The gradient of the perturbation (the `gradient of the gradient' of the temperature) is essentially constant until several hundred metres depth, precluding any measurement of it by InSight.
%AH changed the following sentence
Can any information be gained from the decay with time of a historical heat pulse? We found that the change in heat flow over one simulation time--step (roughly 3000 years) is $\sim1\%$, so only a small fraction of a percent per year. Considering InSight's nominal mission length, this is far below the ability of HP$^{3}$ to detect. \citet{Lorenz2015} likewise showed that the Little Ice Age signature would not show any gradient at the depths to be probed by InSight.

Mean surface temperature is inversely correlated with obliquity \citep{Haberle2003, Forget13}, so the most recent phase of low obliquity, 40 kyr ago, will have led to a small positive excess heatwave, in the same direction as our atmospheric collapse signal. On the other hand, the Little Ice Age would have induced a small negative temperature perturbation. Thus, the two effects might cancel each-other out; or, a suspicious, otherwise unaccounted-for bias in the InSight results, in one direction or the other, might hint at which past climate effect dominates.

If we cannot definitively distinguish the climate change effects, then they will have to be treated as another source of uncertainty, to be mixed in with the instrument error. The total uncertainty would then be $\sim\sqrt{4\%^{2} + 5\%^{2} + 5\%^{2}}\sim 8\%$ (for Little Ice Age and atmospheric collapse signals of $5\%$), which would set the minimum HP$^{3}$ error level and could not be improved upon with deeper or longer measurements.

\section{Conclusion}
\label{S:6}
We estimated the effects on the present--day Martian heat flow of regolith thermal conductivity changes associated with past periods of putative atmospheric collapse, and put them into the context of the InSight heat--flow experiment HP$^3$. During these periods, obliquity changes lead atmospheric CO$_{2}$ to condense out onto the Martian poles, drastically lowering atmospheric pressure and reducing regolith thermal conductivity. Changes in the subsurface temperature profile, as it adjusts to varying thermal conductivity, can then persist for tens of thousands of years and perturb the present--day apparent heat flow.

We used a simple numerical model to calculate the build-up and diffusion of a heat pulse from a recent (40 kyr ago) atmospheric collapse, varying the duration and severity of the collapse, and the depth of the regolith layer to quantify their effects. For $5-10$ m thick regolith layers and the most likely collapse severity, we find subsurface temperatures perturbed by less than 10 K at the end of the collapse which, when evolved forwards with time, corresponds to perturbations of $\sim2-8\%$ of present--day heat flow. This is the positive excess we would expect to measure, on--top of an average geothermal flux of $F=20$ mW m$^{-2}$.

InSight should measure Martian heat flow to an accuracy of between $5$ and $15\%$ (with an instrumental limit of around $4\%$) with its HP$^{3}$ instrument. Thus, in the strongest--case scenario (of full instrument deployment, a complete measurement campaign, and a well-modelled surface configuration), the effects of past atmospheric collapse may be encountered as a slight increase of a few percent in present--day heat flow. Nonetheless, distinguishing this increase from a higher than expected average flux, as well as other climactic changes such as a Little Ice Age, will be challenging, unless a heat--flow measurement can accurately monitor subsurface temperatures for several 100 years and/or measure several 100 m below the surface. As also noted by \citet{Lorenz2015}, the full characterisation of long time-scale perturbations to subsurface temperature requires extremely deep measurements, to depths similar to their skin depths of tens to hundreds of metres, which remains out of the scope of current or near future space missions.

Thus, past climate changes will only be significant as a relatively minor source of added uncertainty in the HP$^{3}$ measurement. If InSight's measurement accuracy proves better than expected (and the other sources of error, such as surface albedo changes, shadowing and re-radiation from the lander, are well understood) then a more detailed model of past climate, incorporating surface temperature trends and gradual rather than instantaneous changes in thermal conductivity, and using in situ determined conductivity, could be considered to further constrain the estimated heat flow perturbation.

\section*{Acknowledgements}
This work was funded by the UK Space Agency, grant no. ST/R001375/2, with additional support from STFC under grant no. ST/S001271/1.

\section*{References} 
\bibliographystyle{model2-names.bst}
\bibliography{sample.bib}

\end{document}